\documentclass[journal]{vgtc}                     

\onlineid{1565}

\vgtccategory{Research}

\vgtcpapertype{Theory/Model}

\title{Data Visualization Style Guides in Practice:\\ Why They Emerge, How They Work, and When They Bend}

\author{%
  \authororcid{Alvitta Ottley}{0000-0002-9485-276X} and 
  \authororcid{Jonathan Schwabish}{0000-0002-3992-6033}
}

\authorfooter{
  \item
  	Alvitta Ottley, Washington University in St. Louis
  	E-mail: alvitta@wustl.edu
  \item
  	Jonathan Schwabish, PolicyViz
  	E-mail: jschwabish@gmail.com
}

\abstract{%
Visualization style guides play a crucial role in shaping how data is interpreted and trusted, yet they often receive little scrutiny in their creation and use. Understanding their impact requires looking beyond the specific rules that style guides prescribe and examining how they function within organizations to coordinate visual work, manage trade-offs, and support judgment under real constraints. 
Analyzing interviews with nine authors of twenty-six style guides across journalism, government, industry, and the public sector, we reveal how these guides reflect the specific challenges of their organizations, including consistency, training, governance, and accountability. Our study highlights the tensions between standardization and flexibility, guidance and discretion, and automation and human oversight.
We propose \acrob{PRISM}, a socio-technical framework that characterizes visualization style guides by their \purpose, \rules, \enforcers, and \flexibility. We show that publicly available style guides expose only a subset of this system.
By viewing style guides as socio-technical systems, we enrich the research on design guidance and offer practical insights for those who create and rely on these guides in critical environments.
}

\keywords{Data Visualization, Style Guides, Design Systems, Interview Study}

\teaser{
  \centering

\includegraphics[width=\linewidth]{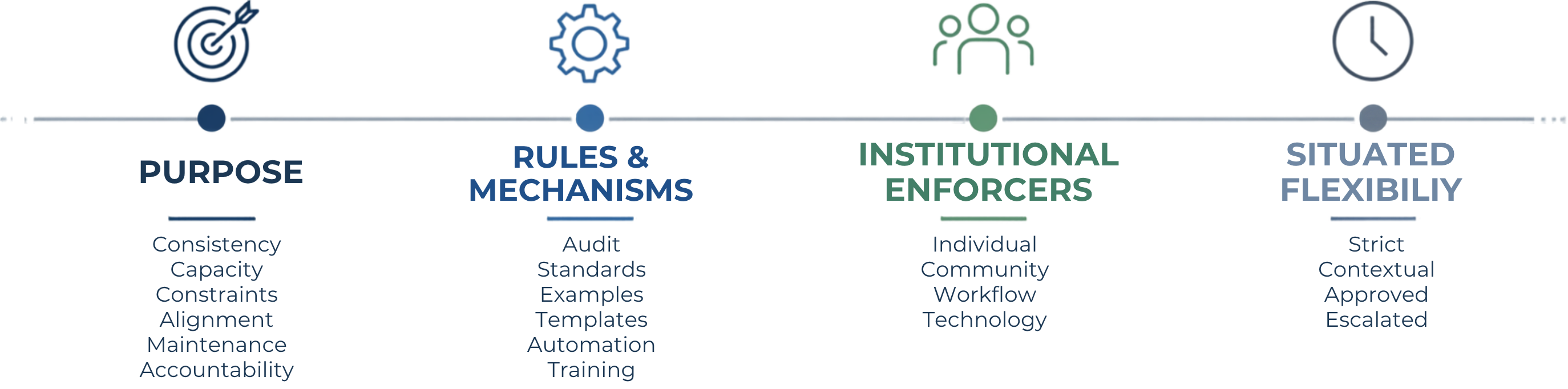}
\caption{An overview of the \acrob{PRISM} framework used to characterize the \acrob{P}urpose, \acrob{R}ule \& mechanisms, \acrob{I}nstitutional enforcers, and the \acrob{S}ituated flexibility of visualization style guides and their \acrob{M}odulation over time.}
\label{fig:teaser-why-how-who-when}

}

\graphicspath{{figs/}{figures/}{pictures/}{images/}{./}} 

\usepackage{tabu}                      
\usepackage{booktabs}                  
\usepackage{lipsum}                    
\usepackage{mwe}                       
\usepackage{multicol}
\usepackage{multirow}
\usepackage{array}
\usepackage{xcolor}
\usepackage{setspace}
\usepackage{soul}

\definecolor{nytInk}{HTML}{111111}
\definecolor{nytGray}{HTML}{6B6B6B}
\definecolor{nytLine}{HTML}{D8D8D8}
\definecolor{nytBG}{HTML}{FFFFFF}
\definecolor{nytWHY}{HTML}{F2F2F2} 
\definecolor{nytHOW}{HTML}{f3f4fb} 
\definecolor{nytWHO}{HTML}{f3fbf5} 
\definecolor{nytWHEN}{HTML}{fbf8f3} 
\definecolor{nytAccent}{HTML}{2A5CAA} 

\newcommand{\palpha}{{\itshape Alpha — Graphics Director, Journalism (US) }}
\newcommand{\pbeta}{{\itshape Beta — Senior Economist, U.S. Government (US) }}
\newcommand{\pgamma}{{\itshape Gamma — Data Visualization Lead, U.K. Local Government (UK) }}
\newcommand{\pdelta}{{\itshape Delta — Data Visualization Designer, Freelance / Consulting (US) }}
\newcommand{\pepsilon}{{\itshape Epsilon — Information Designer, Freelance / Consulting (US) }}
\newcommand{\pzeta}{{\itshape Zeta — Product Design Lead, Technology (US) }}
\newcommand{\peta}{{\itshape Eta — Product Specialist, Visualization Software (DE) }}
\newcommand{\ptheta}{{\itshape Theta — Data Visualization Designer, Freelance / Consulting (DE) }}
\newcommand{\piota}{{\itshape Iota — Data and Visual Journalist, Finance (PT) }}

\usepackage[strict]{changepage}

\usepackage{framed}

\newenvironment{formal}{%
  \MakeFramed{\advance\hsize-\width\FrameRestore}%
  \noindent\hspace{-4.55pt}%
  \begin{adjustwidth}{}{6pt}%
  
  \scriptsize               
  \setstretch{1.0}     

}
{%
  \par 
  \vspace{2pt}\end{adjustwidth}\endMakeFramed%
}

\newcommand{\new}[1]{\textcolor{black}{#1}}

\usepackage{tabularx}

\usepackage{enumitem}
\usepackage{multirow}
\usepackage{microtype}

\usepackage[most]{tcolorbox}

\usepackage{tikz}
\usetikzlibrary{positioning,fit,calc}

\tcbset{
  pigjbox/.style={
    enhanced,
    colback=black!2,
    colframe=black!15,
    boxrule=0.4pt,
    arc=2pt,
    left=6pt,right=6pt,top=3pt,bottom=3pt,
    boxsep=2pt,
    fonttitle=\bfseries,
    coltitle=black,
    attach boxed title to top left={xshift=4pt,yshift=-2pt},
    boxed title style={colback=black!12,colframe=black!12,boxrule=0pt,arc=2pt},
  }
}

\newlist{tightdesc}{description}{1}
\setlist[tightdesc]{
  leftmargin=1.8em,
  labelsep=0.6em,
  itemsep=2pt,
  topsep=2pt,
  parsep=0pt
}

\newcommand{\yes}{\includegraphics[width=11pt]{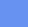}}
\newcommand{\no}{\includegraphics[width=11pt]{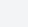}}

\newcommand{\acro}[1]{\textsc{\MakeLowercase{#1}}}
\newcommand{\acrob}[1]{\textsc{\MakeLowercase{\textbf{#1}}}}

\newcommand{\framework}{
\acrob{PRISM}}
\newcommand{\purpose}{
\acro{Purpose}}
\newcommand{\enforcers}{
\acro{Institutional Enforcers}}
\newcommand{\rules}{
\acro{Rules \& Mechanisms}}
\newcommand{\flexibility}{
\acro{Situated Flexibility}}

\newtcbox{\dtag}{on line,
  colback=white,
  coltext=nytGray,
  fontupper=\footnotesize\bfseries\scshape,
  colframe=nytLine,
  boxrule=0.5pt,
  arc=3pt,
  left=1pt,right=1pt,top=1pt,bottom=1pt,
  boxsep=0pt
}

\newtcbox{\pill}{on line,
  colback=white,
  coltext=nytGray,
  fontupper=\footnotesize\bfseries\scshape,
  colframe=nytLine,
  boxrule=0.5pt,
  arc=5pt,
  left=3pt,right=3pt,top=3pt,bottom=3pt,
  boxsep=0pt
}

\newcommand{\PillColW}{2.10cm}
\newcommand{\PillGutter}{0.20cm}
\newcommand{\DescColW}{5.80cm}
\newcommand{\RowGap}{0.01cm}      

\newcommand{\pillrow}[2]{%
  \begin{tabular}{@{}p{\PillColW}@{\hspace{\PillGutter}}p{\DescColW}@{}}
    \hfill\textcolor{nytGray}{\acrob{#1}} & {\footnotesize\textcolor{nytGray}{#2}}
  \end{tabular}\vspace{\RowGap}%
}

\newcommand{\PillColWWhy}{2.60cm}
\newcommand{\PillGutterWhy}{0.20cm}
\newcommand{\DescColWWhy}{5.20cm}
\newcommand{\RowGapWhy}{0.01cm}      

\newcommand{\pillrowwhy}[2]{%
  \begin{tabular}{@{}p{\PillColWWhy}@{\hspace{\PillGutterWhy}}p{\DescColWWhy}@{}}
    \hfill\textcolor{nytGray}{\acrob{#1}} & {\footnotesize\textcolor{nytGray}{#2}}
  \end{tabular}\vspace{\RowGapWhy}%
}

\newcommand{\DimensionCard}[6]{%
\begin{centering}
\smallskip
\begin{tikzpicture}

\def\CardW{\linewidth}

\def\CardH{4.4cm}
\def\mode{#1}%

\def\LongToken{long}%
\def\ShortToken{short}%

\ifx\mode\LongToken
  \def\CardH{4.2cm}%
\else\ifx\mode\ShortToken
  \def\CardH{3.3cm}%
\fi\fi

\def\PadL{0.35cm}
\def\PadR{0.25cm}
\def\PadT{0.35cm}
\def\PadB{0.35cm}

\def\BarInsetX{0.20cm}
\def\BarW{0.08cm}
\def\BarInsetY{0.30cm}

\node[
  draw=nytLine,
  rounded corners=8pt,
  fill=nytWHY, 
  opacity=0.4,
  minimum width=\CardW,
  minimum height=\CardH,
  anchor=north west,
  inner sep=0pt
] (card) at (0,0) {};

\fill[nytAccent]
  ($(card.north west)+(\BarInsetX,-\BarInsetY)$)
  rectangle
  ($(card.south west)+(\BarInsetX+\BarW,\BarInsetY)$);

\coordinate (ct) at ($(card.north west)+(\PadL,-\PadT)$);

\node[anchor=north west, font=\bfseries\Large\scshape, text=nytAccent] (dim)
  at (ct) {\acrob{#2}};

\node[anchor=north west, font=\itshape\footnotesize, text=nytGray]
  at ($(ct)+(0,-0.45cm)$) {#4};

\node[anchor=north west]
  at ($(ct)+(0,-1.0cm)$) {%
    \begin{minipage}[t]{\dimexpr\CardW-\PadL-\PadR\relax}
      #5
    \end{minipage}
  };

\end{tikzpicture}%
\end{centering}
}

\usepackage{mathptmx}                  

\begin{document}


\firstsection{Introduction}

\maketitle


When a visualization looks ``wrong,'' it is often easy to diagnose potential issues. This is in part because visualization research has made substantial progress in identifying design principles and best practices that help explain when and why visualizations fail. 
When a visualization looks ``right,'' however, far fewer people ask why or how particular design choices come to feel natural, credible, or appropriate in the first place. These judgments are rarely accidental.
Behind many of the seemingly intuitive design choices we encounter in journalism and from organizations with established data visualization design teams lie \textit{visualization style guides} or \textit{design systems}~\cite{ottley2026consensus,dvsg2026,elder2020should}. These are artifacts that consolidate learned knowledge, quietly shaping how data is seen, interpreted, and trusted at scale. Despite their growing influence across journalism, government, industry, and the public sector, style guides remain largely absent from visualization research. As a result, we know surprisingly little about how these guides are created, interpreted, and enforced.

Existing analyses of style guides rely primarily on publicly available artifacts, which emphasize advice, rules, examples, and templates (e.g., ~\cite{ottley2026consensus, kandogan2016grounded, choi2021toward}). However, these artifacts provide only a partial view. Arguably, the most consequential aspects of style guides, i.e., their motivations, governance, and flexibility, are often implicit and embedded in organizational practice.
Thus, existing work provides rich accounts of what visualization guidelines contain and how they might be formalized~\cite{ottley2026consensus}. However, less attention has been paid to how such guidance is created, maintained, and enacted within organizations.

To understand how visualization style guides actually operate in practice, we conducted an interview study with authors and maintainers of data visualization style guides across diverse domains, including journalism, government, industry, and the public sector. Many participants had developed multiple guides over extended periods and across organizations, offering longitudinal and comparative perspectives. Rather than focusing on the content of individual rules, we asked how guides emerge, how they are adopted, how they change over time, and how practitioners decide when to follow or break them.

Our analysis reveals that style guides consistently emerged in response to organizational pressures (e.g., design inconsistencies or branding). These pressures motivated concrete interventions such as templates, checklists, tool defaults, and exemplar libraries. However, participants emphasized that codified rules alone were insufficient. As visualization work expanded across people, platforms, and contexts, effective guides depended on governance mechanisms (e.g., champions, guardians, working groups, tool integration) and calibrated flexibility regimes (e.g., strict, contextual, approved, escalated) that structured when and how deviation was permitted.

Synthesizing these findings, we introduce \acrob{PRISM}, a four-dimensional framework that characterizes visualization style guides along interacting axes: Purpose, Rules \& Mechanisms, Institutional Enforcers, and Situated Flexibility. We then analyzed 50 publicly available style guides to examine what aspects of guidance are observable in documentation.

\acrob{PRISM} reframes style guides as sociotechnical systems or solutions to particular organizational problems that combine rules, tools, governance, and judgment. By shifting the unit of analysis from isolated guidelines to the organizational ecosystems in which they are produced and enacted, we offer a broader account of how ``what looks right'' becomes stabilized in practice. We make four primary contributions:
\begin{itemize}
    \item \textbf{An empirical account of visualization style guides.} We provide a cross-domain account of how style guides are initiated, structured, maintained, and adapted through interviews with authors and maintainers.

    \item \textbf{\acrob{PRISM}, a socio-technical framework for understanding style guides.} We introduce a conceptual framework organized around the Purpose, Rules, \& Mechanisms, Institutional Enforcers, and Situated Flexibility that shape style guide practice.
    
    \item \textbf{Revealing the partial observability of style guides.} We analyze 50 publicly available documents, and we show that these artifacts primarily expose rules and mechanisms, while purpose, enforcement structures, and situational flexibility are often implicit.

\end{itemize}
\begin{table*}[t!]

\caption{The interview participants, their professional roles, sectors, experience, and domains in which they authored or maintained style guides.}
\vspace{-.2cm}
\centering
\small
\begin{tabular}{l l l l c l c l}
\toprule
\textbf{Participant} & \textbf{Locale} & \textbf{Role} &  \textbf{Sector} & \textbf{\# Guides} & \textbf{Select Domain Coverage} & \textbf{Experience} & \textbf{Maintenance} \\
\midrule
Alpha   
& US
& Graphics Director 
& Journalism
& 3 
& Journalism, Education, Policy Institute 
& 20+ yrs 
& -- \\

Beta    
& US
& Senior Economist 
& U.S. Government 
& 1 
& Government (Public Health) 
& 10+ yrs 
& -- \\

Gamma 
& UK
& Data Visualisation Lead 
& U.K. Local Government 
& 1 
& Government (Urban Services) 
& 7+ yrs  
& In progress \\

Delta 
& US
& Data Visualization Designer 
& Freelance / Consulting 
& 7 
& Foundations, Business Intelligence 
& 12+ yrs 
& -- \\

Epsilon 
& US
& Information Designer 
& Freelance / Consulting 
& 5 
& Accounting, Pharmaceutical, FinTech 
& 8+ yrs 
& -- \\

Zeta 
& US
& Product Design Lead 
& Technology 
& 1 
& Technology (Software)
& 15+ yrs 
& In progress (6 updates) \\

Eta  
& DE
& Product Specialist 
& Visualization Software 
& 1 
& Policy Institute 
& 7+ yrs 
& -- \\

Theta
& DE
& Data Visualization Designer 
& Freelance / Consulting 
& 5 
& Transportation, Intergovernmental 
& 8+ yrs 
& 1 year \\

Iota
& PT
& Data and Visual Journalist 
& Finance 
& 2 
& Journalism, Finance 
& 4+ yrs 
& In Progress \\

\bottomrule
\end{tabular}

\vspace{-.3cm}
\label{tab:participants}
\end{table*}

\section{Background}

Visualization research has long sought to articulate principles that improve how data is perceived, interpreted, and acted upon. Foundational work on graphical perception~\cite{cleveland1984graphical}, visual encodings~\cite{bertin1967semiologie}, information mapping~\cite{card2009information}, and interaction design~\cite{shneiderman2003eyes} established systematic approaches to designing effective visual representations. Subsequent research has expanded this foundation to examine how cognitive, narrative, ethical, and contextual factors influence comprehension, trust, and decision-making~\cite{hullman2011visualization,correll2019ethical,lin2025makes,mckinley2025trustworthy,pandey2023you}. Collectively, this work provides a rich theoretical and empirical basis for visualization guidance.


\subsection{From Visualization Principles to Style Guides}

Style guides are one mechanism through which visualization principles are translated into shared practice. Prior research has examined such guidance from multiple perspectives. Academic survey papers synthesize empirical findings into consolidated recommendations (e.g., \cite{franconeri2021science,zeng2023review}), while historical analyses trace graphic rules to early works such as Brinton’s \textit{Graphic Methods for Presenting Facts}~\cite{brinton1919graphic,elder2020should}.

More recent work has analyzed how guidelines are structured, categorized, and debated in practice. Kandogan and Lee~\cite{kandogan2016grounded} examined hundreds of guidelines from academic and practitioner materials, showing that guidance spans data characteristics, analytic tasks, user expertise, and insight, and not only visual form. Choi et al.~\cite{choi2021toward} decomposed guidelines into structural components and highlighted opportunities for formalization and automation. Diehl et al.~\cite{diehl2020studying} studied how practitioners invoke guidelines in online forums, finding that guidance is frequently referenced but inconsistently applied. Corpus-based analyses have compared recommendations across organizations, revealing convergence, contradiction, and embedded institutional values~\cite{ottley2026consensus}.


\subsection{From Prescriptions to Practice}

Visualization guidance has been operationalized along two primary trajectories. First, visualization authoring systems, such as Vega-Lite~\cite{satyanarayan2016vega}, CompassQL~\cite{wongsuphasawat2016towards}, and Draco~\cite{moritz2018formalizing}, encode design heuristics directly into declarative grammars and recommendation engines. Many of these systems originated as research projects and therefore provide explicit documentation of their assumptions, constraints, and rationale.

Second, organizations translate accumulated research findings and practitioner norms into visualization \textit{style guides} or \textit{design systems}: structured collections of rules, templates, and defaults intended to promote consistency and quality. Unlike research-driven systems, their underlying rationales are often implicit rather than formally articulated.

Moreover, recent scholarship has emphasized sociotechnical perspectives, including value-sensitive design~\cite{friedman1996value,friedman2017survey}, ethical reflection~\cite{correll2019ethical,diakopoulos2018ethics}, and the role of institutional context in shaping analytic systems~\cite{wang2024card}. These perspectives suggest that design conventions may encode organizational priorities and social values in addition to perceptual and cognitive efficiency.\smallskip

\noindent
In sum, prior work relies on visible artifacts such as rules, examples, and templates, creating a structural bias toward what is documentable and underexplores the organizational and contextual dimensions of guidance. In contrast, our work shifts the unit of analysis from guidelines as artifacts to style guides as socio-technical systems.

\section{Methodology}

Many aspects of style guide practice, such as why guidance was created, how it is enforced, and when it is adapted, are not captured in public documentation. 
Our goal was to understand these unobservable aspects. 
Because these processes are shaped by tacit knowledge, organizational structure, and situated judgment, we employed a qualitative interview methodology centered on practitioners with direct experience authoring or maintaining visualization style guides.

\subsection{Participants}

We conducted semi-structured interviews with nine authors and maintainers of data visualization style guides from diverse domains, including journalism, government, industry, and non-profit organizations. Participants were recruited through professional networks, prior collaborations, and public documentation of visualization style guides. We intentionally sampled participants with substantial experience producing or maintaining style guides, including several who had developed multiple guides across different organizations or over extended periods. 
Our goal was not statistical generalization, but to identify recurring structural patterns across diverse settings.
Interviews were conducted until additional participants reinforced existing patterns. 
Table~\ref{tab:participants} summarizes the experience of participants in this study.

\subsection{Interviews}
Interviews were conversational in structure, allowing participants to surface unanticipated issues and reflect on tradeoffs they encountered. \new{
Seven of the nine interviews were conducted jointly by both authors. The initial interviews were led by Schwabish, who has professional interviewing experience, to establish consistency and refine the protocol. As the protocol stabilized, interviewing responsibilities alternated between the authors. The remaining two interviews were conducted individually, one by each author. Conducting interviews jointly also enabled immediate discussion of emerging themes.}

Interviews were conducted remotely via Zoom, lasted approximately 60 minutes, and were audio-recorded with participant consent. Recordings were transcribed verbatim for analysis. Core topics included: motivations for creating a guide; organizational context and constraints; early design decisions; adoption and resistance; tooling and infrastructure; maintenance and governance; and situations in which rules were followed, overridden, or negotiated.  We provide the complete list of questions in Appendix~\ref{app:interview-protocol}. To protect participant privacy and organizational confidentiality, we anonymized transcripts and refer to participants using pseudonyms. This study was reviewed and approved with an exempt status under Washington University's \acro{IRB} protocol.

\vspace{2mm}

\subsection{Analysis Approach}

We employed reflexive thematic analysis (RTA), a qualitative approach introduced by Braun and Clarke~\cite{braun2019reflecting,braun2021one,braun2023doing,campbell2021reflexive} for identifying patterns of shared meaning through iterative and interpretive analysis. 
Unlike coding approaches that emphasize categorization consistency or frequency counts, \acro{RTA} treats themes as analytic constructs that are actively developed and refined by the researcher to capture underlying mechanisms and relationships within the data. This is particularly important in our context, where practices vary across organizations but share underlying functional roles.
\acro{RTA} follows a six-phase iterative process: 

\begin{enumerate}
    \item \textbf{Familiarization:} We conducted multiple close readings of each transcript to identify concrete practices, organizational constraints, and decision-making tensions within and across interviews.
    
    \item \textbf{Coding:} Initial codes captured both concrete practices (e.g., need analysis, template development) and inferred dynamics (e.g., governance practice, use flexibility).

    \item \textbf{Generating Initial Themes:} Style guides were often described as responses to organizational \acro{problems}, while their development and implementation took the form of corresponding \acro{intervention} strategies. Additional themes captured variations in \acro{governance} structures (i.e., who enforces guidance) and mechanisms for \acro{flexibility} or exception handling, foreshadowing the four-dimensional structure developed in later phases.

    \item  \textbf{Reviewing Themes:} Candidate themes were then evaluated for coherence and distinctiveness. For example, we initially considered elevating ``Diagnostics'' as a distinct theme to capture needs analysis activities. However, these activities were consistently described as preparatory steps embedded within solution development. We therefore reconceptualized them within a broader intervention category (e.g., \acro{Audit}), to reduce redundancy and clarify the functional relationships between themes.

    \item \textbf{Defining and Naming Themes:} Higher-level constructs were progressively abstracted and consolidated into the four interacting dimensions introduced in the next section: \purpose\ (\textit{the why}), \rules\ (\textit{the how}), \enforcers\ (\textit{the who}), and \flexibility\ (\textit{the when}). 

    \item  \textbf{Write-Up and Validation} We refined the final framework during the writing process by continually referencing transcript excerpts. We actively searched for disconfirming cases and examined available style guide documents and tooling artifacts to assess alignment between reported practices and material implementations. Interpretations were revised where ambiguities emerged. 
\end{enumerate}

\new{Because both authors were present during most interviews, candidate themes were documented independently as they emerged. These observations were then compared and consolidated through discussion, providing an initial analytical foundation before formal coding. Ottley subsequently led transcript familiarization, coding, and theme refinement, while both authors regularly reviewed interpretations, challenged emerging themes, and refined the developing framework. Consistent with reflexive thematic analysis, our goal was not coder agreement but the iterative development and refinement of themes.}

We initially considered common methodologies for quantitative analysis, such as independent coding and inter-coder reliability. However, counts derived from interview data can be misleading. Given the conversational nature of the interviews and the diversity of participants’ roles and experiences, the absence of a theme in a given transcript does not necessarily indicate its absence within an organization. It may reflect that it did not arise in that particular discussion or fell outside the participant’s direct experience. Appendix~\ref{app:participant-coverage} reports participant-level coverage of major framework attributes. However, it should not be interpreted as prevalence.

\section{\acrob{PRISM} Framework Overview}

Synthesizing findings, we introduce \acrob{PRISM}, a socio-technical framework for analyzing visualization style guides as organizational artifacts.
The framework is organized around four themes that participants repeatedly surfaced, sometimes implicitly, when describing their experiences authoring, maintaining, and using style guides:

\begin{itemize}
  \item \purpose: The pressures and constraints motivating the creation. 
  \item \rules: The mechanisms through which guidance is expressed and operationalized.
  \item \enforcers: The actors and structures responsible for maintaining and applying guidance.
  \item \flexibility: The conditions under which guidance is enforced, adapted, or overridden.
\end{itemize}


Importantly, these dimensions interact in systematic ways. For example, high-accountability contexts often lead to stricter enforcement regimes and greater reliance on workflow or technological enforcement. Similarly, capacity constraints frequently drive the adoption of templates and automation, reducing reliance on individual expertise. These patterns suggest that style guides are not modular collections of rules, but tightly coupled configurations shaped by organizational priorities.

\DimensionCard{long}{Purpose}{}{Why was the style guide created?}{
  \pillrowwhy{consistency}{reduce variation across teams and outputs}
  \pillrowwhy{alignment}{ensure visualizations reflect the brand identity}
  \pillrowwhy{capacity}{reduce reliance on scarce expertise}
  \pillrowwhy{maintenance}{keep guidance current as needs evolve}
  \pillrowwhy{constraints}{address platform limits and set default}
  \pillrowwhy{accountability}{manage risk, scrutiny, and accountability}
}


\section{Purposes and Institutional Pressures}


If style guides were merely collections of ``best practices,'' we would expect their motivations to be largely aesthetic and their contents to converge. Our interviews reveal a broader picture. Data visualization style guides are often authored in response to breakdowns that arise when visualization work is produced by many people, across many tools, or under public scrutiny. These problems are not limited to visual inconsistency or aesthetic drift; they also include challenges of coordination, \acro{capacity}, technological \acro{constraints}, \acro{maintenance}, and \acro{accountability}, many of which are invisible when style guides are viewed only as design rules. Participants repeatedly described creating guidance without a clear playbook for how to do so, often beginning with personal frustration or an acute organizational need.

\subsection{Consistency and Reducing Variation}
Consistency was one of the most widely cited motivations, and participants described it less as a preference for uniformity and more as a response to fragmentation. Many recounted observing that charts produced by different people (or in different tools) varied in typography, color, layout, and even vocabulary, making outputs feel incoherent even within the same department.

\begin{formal}
``I remember we were all using Excel spreadsheets, and all our charts \ldots they looked totally different from each other.''

\hfill
\peta
\end{formal}

Participants often framed consistency as making a unit’s output ``feel like us'' 
and reducing the cognitive burden of re-learning a visual language from chart to chart. Several described concrete catalysts, often variations in \textit{color}, as the most visible symptom of inconsistency and the easiest entry point for intervention.

\subsection{Alignment and Brand Identity}
Although brand identity was sometimes cited as a key motive for the style guide, participants did not describe alignment solely as ``branding.'' Instead, our framework uses alignment to encompass a broader effort to ensure that visualization outputs are recognizable as coming from the organization and consistent with its values, such as clarity, accessibility, and, in some cases, ethics. Practical triggers included company rebrands and organizational shifts that required rethinking color palettes and typographic choices, as well as ongoing negotiation with upstream brand teams and design systems.

\begin{formal}
``They rebranded because they wanted to be quieter\ldots complete reverse… it had a huge impact\ldots the type was quieter, and it changed the type of charts we're making.''

\hfill
\pepsilon
\end{formal}

Alignment also reflected audience positioning. Some organizations prioritized distinctiveness over accessibility, while others treated ethics and inclusion as first-order design values. In several accounts, alignment therefore describes an \emph{organizational stance} i.e., a decision about what the organization wants its visualizations to signal, whom they are for, and which tradeoffs are acceptable.

\subsection{Capacity, Limited Time and Expertise}
Capacity pressures were pervasive and multifaceted. Participants described repeated cycles of correction, long revision loops, difficulty onboarding new members due to unwritten rules, and the practical reality that many people producing charts are not trained visualization designers. In these contexts, guides were motivated by the need to reduce repeated decision-making and to help non-experts create ``good enough'' work without continuous expert intervention.

\begin{formal}
``We're, I don't know, 800 people right now at [redacted], and our graphics team has\ldots 4 or 5 different editors, like, that's it. Like, we don't have the bandwidth to be able to micromanage and edit every single chart.

\hfill
\palpha
\end{formal}

Participants emphasized that capacity is not only about speed but also about organizational sustainability: inefficiency leads to late nights, stalled projects, and an inability for experts to focus on higher-value work. Several accounts also cautioned that capacity is shaped by \emph{adoption friction}. If guidance is hard to find, too long to read, or unsupported by training and feedback loops, it fails to relieve the underlying burden.

\subsection{Constraints and defining system defaults.}
Several participants emphasized that style guidance becomes actionable only when it can be \emph{carried by the production environment}. Instead of treating tools as external constraints to work around, many described style guides as a translation layer that converts guidance into defaults, templates, tokens, and components that fit within (and sometimes are introduced alongside) a particular tool or design system. In these accounts, it was important that the tools do not limit what is possible, but defaults are warranted, especially when distributed teams cannot reliably remember guidance unless it is embedded where work happens.

\begin{formal}
``We were moving from our old design system \ldots And so we need to decide what do [redacted] charts look like.''

\hfill
\pzeta
\end{formal}

This embedding work was triggered by platform transitions (e.g., moving to a new design system) or by the need to support multiple environments (web, mobile, desktop) with consistent behavior. Participants described this as a recurring struggle: tools are rarely ``opinionated'' about data visualization, and organizations therefore invest in constructing opinionated defaults to avoid re-deciding foundational choices across projects. Importantly, this category is distinct from maintenance. \acro{Constraints} describe the need to operationalize guidance within a particular design environment, whereas \acro{maintenance} describes the ongoing work required to keep those embedded defaults aligned as tools, requirements, and organizational priorities change.

\subsection{Maintenance, Updates, and Evolution.}
Participants repeatedly characterized style guides as ``living'' systems that require ongoing maintenance, yet some also described the lack of funding, time, and mechanisms to sustain that work. Updates were triggered by rebrands, platform redesigns, team growth, and the realization that earlier guidance had become outdated or incomplete.

\begin{formal}
``Style guides are living documents; they need to grow and evolve.''

\hfill
\pdelta
\end{formal}

Conversations about updates also revealed a common pattern of \emph{underestimation}. Some teams initially assumed revision would be quick, only to discover that doing it well required revisiting fundamental decisions and anticipating future constraints, an effort measured in months or years.

\subsection{Accountability and responsibility}
A striking observation was how ``looking right'' is inseparable from \emph{being defensible}. In government and regulated domains, accountability was described in legalistic terms. Organizations face formal requirements and the prospect of being challenged or sued. 
During our initial conversations, \textit{Theta}, a consultant for an intergovernmental organization, explained that the organization's decision to develop a design system, including style specifications, was motivated in part by concerns surrounding accessibility, internationalization, and equity.

\begin{formal}
``I realized the [redacted] does not do a good job in all these areas, which is absurd, because it should be the one organization that does an excellent job.''

\hfill
\ptheta
\end{formal}




Accountability was also expressed as a reputational risk and a financial consequence. Participants working with business and finance organizations linked visualization quality to how boards, investors, customers, and shareholders interpret credibility and to the possibility that flawed or misleading charts would be read as manipulation.

\begin{formal}
``If you have bad charts, there will be consequences \ldots how much money you raise, how does your board feel about things, how does the VC \ldots how does your customer feel?''

\hfill
\pepsilon
\end{formal}

In public-sector intelligence and public health contexts, accountability was framed as the need for clarity and restraint with data communication because the stakes are immediate and consequential.

\begin{formal}
``Now I can't play quite as fast and loose in the data arts space, because \ldots this stuff matters \ldots we can't just express freely here, this needs to be rigid.''

\hfill
\pgamma
\end{formal}

These accounts suggest that accountability is fundamentally different from other motivations, such as aesthetics or efficiency. It is the \emph{contextual frame} that shapes how other pressures are interpreted. When the risk of misinterpretation is high, organizations seek guidance that is more explicit, less negotiable, and more enforceable.

\subsection{Synthesis}
These pressures explain why style guides vary dramatically in structure, format, and content across organizations. A guide motivated primarily by \acro{Consistency} and \acro{Capacity} may emphasize templates, restricted palettes, and a small set of chart types. A guide motivated by \acro{Alignment} may foreground brand values, typographic systems, and tone. A guide operating under \acro{Accountability} may become more rigid, explicit, and auditable. In other words, style guides are not merely inventories of rules; they are problem-specific organizational solutions whose form reflects the pressures they are designed to relieve.

\DimensionCard{long}{Rules \& Mechanisms}{}{How is guidance expressed or operationalized?}{
  \pillrow{audit}{review outputs identify issues \& patterns}
  \pillrow{standards}{explicit rules for visual design decisions}
  \pillrow{examples}{create exemplars and do/don’t comparisons}
  \pillrow{templates}{embed guidance into reusable structures and defaults}
  \pillrow{automation}{enforce guidance through code or tooling}
  \pillrow{training}{support adoption through onboarding and workshops}
}

\section{Rules and Intervention Mechanisms}

Building on the \purpose\ dimension, which explains the pressures that give rise to style guides, we explore how organizations attempt to intervene. Participants did not describe a single dominant model of guidance. Our interviews revealed a spectrum of mechanisms, ranging from diagnostic audits and narrative examples to highly embedded automation within software systems.
In practice, most organizations combine several of these approaches.

\subsection{Audit: Making the Invisible Visible}

For many participants, particularly consultants, intervention began with an audit. Audits involved systematically reviewing existing charts, documentation, and user workflows to identify recurring patterns and breakdowns, and often standards or existing guidelines from similar organizations.

\begin{formal}
``we always do a big audit first. That's the first step''

\hfill
\textit{Epsilon — Information Designer, Consulting (US)}
\end{formal}

Audits involved multiple processes. For example, some participants described how they cataloged chart types, data types, color usage, and publication formats to ``zoom out'' and see patterns that teams had not previously recognized. Their review and quantitative analysis of visualization artifacts were typically supplemented with interviews and surveys to understand how charts were used and interpreted.

\begin{formal}
``Looking at different charts that they've made in the past, and figuring out what is working, what's not working.''

\smallskip
\noindent
``It's a combination of looking at the data and then, like, talking to people to sort of figure out what that actually means.''

\hfill
\textit{Delta — Visualization Designer, Consulting (US)}
\end{formal}

Thus, these audits served as diagnostic investigations to ground style guides, updates, or design systems in observed practice rather than in abstract ideals.

\subsection{Standards: Writing the Rules Down}
\label{sec:how:standards}

A second, more explicit mechanism involved codifying standards. Participants described selecting and documenting recommendations for typography, color palettes, layout systems, accessibility requirements, naming conventions, and ethical constraints. Standards ranged from high-level principles to highly detailed specifications. Some included explicit accessibility thresholds and contrast ratios; others defined baseline rules, color hierarchies, and even tick mark frequencies for engineering implementation.

\begin{formal}
``Titles should be this number \ldots the intro text should be this.''

\smallskip
\noindent
``there are typography rules, there are color rules, there are, you know, best practices, that kind of stuff.''

\hfill
\palpha
\end{formal}

In several accounts, standards also encompassed ethical guidance:

\begin{formal}
``For them, ethics was actually a primary value,\ldots it was super important that we would consider, like, gender bias, and colors, in treatment, in iconography.''

\hfill
\pepsilon
\end{formal}

Standards transform tacit knowledge into shared reference. However, they also introduce interpretive burden; most participants stressed that documentation alone does not guarantee consistent adoption.


\subsection{Examples: Showing What Good Looks Like}
\label{sec:how:examples}

To reduce ambiguity, many style guides relied heavily on examples. Participants described, including side-by-side comparisons of good and bad charts, annotated diagrams, anatomy breakdowns, and even reference implementations.

\begin{formal}
``We have side-by-side examples of good and bad, went more in-depth in a couple of big ones that really we saw all the time.''

\smallskip
\noindent
``You take this default graph, and here are all the choices that were made to make these improvements, so it made it a little more accessible.''

\hfill
\pbeta
\end{formal}



Examples functioned as translation devices between abstract standards and practical application. Several participants emphasized that showing rather than merely telling was essential for uptake. Additionally, authors frequently reviewed other organizations’ guides, borrowing structures and visual idioms, suggesting that examples circulate across institutional boundaries and shape emerging conventions.

\subsection{Templates: Structured Starting Points}

\acro{Standards} describe and \acro{Examples} demonstrate; several participants discussed how they used templates to operationalize. \acro{Templates} included \textit{Excel} starters, \textit{Tableau} themes, \textit{R} packages, \textit{Figma} components, boilerplate notebooks, and printable one-pagers.

\begin{formal}
``we kind of built out some things in Excel and Tableau as kind of starter places for folks to tap into.''

\hfill
\pbeta
\end{formal}

Templates reduced friction by pre-configuring key decisions. At the same time, participants noted tradeoffs. Overly rigid templates risked encouraging superficial compliance without deeper understanding. In some contexts, templates were viewed as scaffolds, not constraints, and were starting points that could be extended or adapted.

\subsection{Automation: Embedding Rules in Software}

The most restrictive and scalable mechanism involved embedding standards directly into tools and component systems. Here, the style guide becomes infrastructural, and the tool enforces the rules.

\begin{formal}
``We give them the tool, and the tool is the style guide at that point.''

\hfill
\palpha
\end{formal}

Participants described \textit{R} packages with preset themes, design tokens exported into component libraries, and visualization systems where color palettes, grid lines, fonts, and accessibility constraints were baked into defaults. In doing so, automation reduces reliance on individual memory and increases consistency, but often at the cost of flexibility.

\subsection{Training: Teaching Interpretation and Judgment}

Finally, participants emphasized training as essential for adoption. Guides served as curricular foundations for workshops, onboarding sessions, community-of-practice meetings, and hands-on critiques.

\begin{formal}
``we kicked it off with an in-house, day-long workshop.''

\hfill
\pdelta
\end{formal}

Training addressed not only procedural knowledge (e.g., how to apply color, how to choose a chart type) but also conceptual reasoning (e.g., how to think about audience, accessibility, and ethics). Several participants noted that documentation alone is insufficient, and sustained engagement through workshops, critiques, and office hours is needed to reinforce interpretation and community norms.

\subsection{Synthesis}

Organizations rarely relied on a single mechanism. Instead, they layered audits to diagnose problems, standards to codify solutions, examples to illustrate intent, templates to scaffold production, automation to enforce defaults, and training to sustain understanding.
Many guides are available in various document formats, such as \acro{PDF} or \acro{HTML}; however, there is a growing trend toward creating interfaces that are more browsable and searchable. Additionally, several participants highlighted ongoing efforts or expressed interest in using AI tools to automate compliance detection and facilitate direct linting. Thus, the process of creating and enforcing style guides is actively evolving. \par

\DimensionCard{short}{Institutional Enforcers}{}{Who makes the guidance work?}{
  \pillrow{individual}{relies on personal judgment and local expertise}
  \pillrow{community}{maintained through shared norms and peer feedback}
  \pillrow{workflow}{enforced through procedures and review stages}
  \pillrow{technology}{embedded in systems, platforms, and infrastructure}
}

\section{Institutional Enforcers}

Participants made clear that style guides do not govern themselves. Their effectiveness depends on who stewards them, who has decision rights, and whether guidance is upheld socially, procedurally, or infrastructurally. We observed four primary enforcement models: \acro{Individual}, \acro{Community}, \acro{Workflow}, and \acro{Technology}. These modes differ in how authority is exercised, how scalable they are, and how much they rely on trust versus formal structure.

\subsection{Individual: Guardians, Champions, and Experts}

In many organizations, coordination is centered around one or a small number of individuals. These actors functioned as editors, administrators, evangelists, or consultants, holding both expertise and informal authority.
These individuals frequently described themselves, or were described by others, as \textit{champions}, \textit{guardians}, or \textit{experts} whose credibility enabled adoption.

\begin{formal}
``if I wanted \ldots I could go edit the charts''

\hfill
\piota
\end{formal}

Individual coordination offers agility and coherence, particularly in small teams. However, it is fragile as it depends on the continued presence, authority, and bandwidth of specific people. Several participants noted that their ability to ``police'' or approve work was limited to certain teams, highlighting the limitation of individual stewardship.

\subsection{Workflow: Process, Review, and Shared Governance}

Some organizations integrated coordination into their formal workflows. Instead of depending solely on individual authority, they established structured review processes, sign-offs, working groups, and regular meetings to share responsibility.
Additionally, workflow coordination included structured feedback loops, weekly design reviews, workshops, quality and checklists.

\begin{formal}
``we have a weekly meeting when we have a design review\ldots and the team gets feedback.''

\hfill
\piota
\end{formal}

Workflow-based coordination scales more reliably than individual stewardship, but it introduces overhead and depends on sustained participation and institutional buy-in.

\subsection{Community: Coalitions and Distributed Ownership}

A third enforcement model operated through community: networks of practitioners who sustain guidance through shared norms, mutual support, and repeated interaction. Unlike workflow coordination, which is enforced by process, community coordination is sustained by people knowing who else makes charts, having channels to ask questions, and developing a shared sense of what ``good'' looks like.

Participants described deliberately widening involvement beyond a single professional group. This work often began with mapping who produces visualizations and intentionally creating spaces for interaction.

\begin{formal}
``You need to have a working group once a quarter, once a month. Among people at your organization who make charts, because they should know each other to at least ask questions.''

\hfill
\pdelta
\end{formal}

Community-based coordination also relied on a lightweight communication infrastructure (e.g., office hours, email lists, and chat channels) that keeps guidance usable when people encounter edge cases.

\begin{formal}
``we always recommend that they have, like, an email channel, office hours, Slack channel, something like that, where people can go and ask questions when they get stuck.''

\hfill
\pepsilon
\end{formal}

Community coordination was not limited to internal networks. Some participants drew ideas, troubleshooting, and norms from external communities of practice.

\begin{formal}
``There was a community\ldots the Data Visualization Society\ldots by talking to other people, I would have questions… and I would grab ideas.''

\hfill
\peta
\end{formal}

Several participants emphasized that community support is also a mechanism for adoption, expressing that buy-in emerges when people see exemplars and feel both bottom-up and top-down support.
This highlights a distinct form of governance where guidance can be produced \emph{with} users rather than \emph{for} them, through participatory scoping and iterative feedback.
This enables style guides to spread ``organically'' through a coalition of the willing, instead of through mandate.

\subsection{Technology: Infrastructural Enforcement}

In some cases, coordination shifted from people and processes to infrastructure. When standards are embedded in tools, templates, or component systems, the system itself becomes the enforcer.

\begin{formal}
``We're using DataWrapper now\ldots Most simple charts are effectively pre-programmed. Set it and forget it.''

\hfill
\palpha
\end{formal}

Participants described building wrappers, component libraries, and tool extensions to ensure that guidance was ``baked into the software.'' Such technology-based enforcement reduces reliance on memory, persuasion, and review. It scales efficiently and minimizes deviation for routine work. However, it can also constrain flexibility and shift complexity upstream to those responsible for building and maintaining the infrastructure.

\subsection{Synthesis}

These enforcement models are not mutually exclusive. Most organizations combine them: an expert champion might define standards, community channels might sustain day-to-day use, workflow processes might review exceptions, and tools might enforce defaults. Nevertheless, the dominant locus of coordination shapes how guidance is experienced.
Understanding \emph{who} coordinates guidance, therefore, reveals how power, flexibility, and responsibility are distributed within an organization.\\
\par

\DimensionCard{short}{Situated flexibility}{}{When are exceptions allowed and how are they handled?}{
  \pillrow{strict}{rules are applied without exception}
  \pillrow{contextual}{exceptions permitted based on situation or audience}
  \pillrow{approved}{exceptions require explicit sign-off by ``guardian''}
  \pillrow{escalated}{exceptions require higher-level review}
}

\section{Situated Flexibility and Exemption Handling}

This dimension captures how rigidly guidance is applied in practice. Across interviews, participants emphasized that rules are rarely simply followed or ignored. Instead, they are enacted under different conditions of flexibility.
We observed four exception handling models: \acro{Strict}, \acro{Contextual}, \acro{Approved}, and \acro{Escalated}. These models describe the conditions under which deviation is tolerated, negotiated, or prohibited.

\subsection{Strict: Non-Negotiable Requirements}

In some domains, certain standards were described as non-negotiable. These rules applied across contexts, tools, and teams without exception.
Strict enforcement was often justified by risk, brand integrity, or accessibility.

\begin{formal}
``They really, really, really valued accessibility to a standard that I've never seen before\ldots especially with color, it was really apparent that all of their colors had to be a minimum of 3 to 1 contrast against each other.''

\hfill
\pdelta
\end{formal}

Some participants described technological lock-in as a deliberate strategy for maintaining strictness.

\begin{formal}
``Locking you in is a good thing, because when somebody decides they want the title of the chart to be bright red, it doesn't let them do that.''

\hfill
\palpha
\end{formal}

Under a \acro{strict} model, deviations are either technically impossible or socially unacceptable. This model was most common in high-stakes, externally visible, or legally sensitive contexts.

\subsection{Contextual: Situation Dependent}

More commonly, participants described guidance as conditional rather than absolute. Rules provided defaults or starting points, but exceptions were permitted depending on the audience, deadline, medium, or story importance.

\begin{formal}
``We made a very deliberate decision to do almost no enforcement. So we will help, but we will not police.''

\hfill
\peta
\end{formal}

All participants explicitly rejected rigid prescriptions when context demanded variation. Participants frequently emphasized that black-and-white right-or-wrong binaries do not reflect the realities of practice. In the \acro{contextual} model, judgment is distributed to practitioners. Flexibility is intentional, often framed as necessary for storytelling, medium adaptation, or evolving design practice.

\subsection{Approved: Exceptions with Guardian Sign-Off}

In some organizations, deviation from standards was possible but required explicit approval from a designated steward, often an editor, administrator, or ``guideline guardian.'' One participant recalled approving an exception:

\begin{formal}
``\ldots and the justification is because it looks prettier.''

\hfill
\piota
\end{formal}

Here, flexibility is neither fully decentralized nor fully locked in. Instead, exceptions are mediated by expertise. 
Additionally, participants described variations of this model in which headquarters or a central team defined the standard, but adoption elsewhere remained voluntary or negotiated.

\begin{formal}
``this is what we are doing at headquarters, you can adopt this or not, but this is something that will help.''

\hfill
\pbeta
\end{formal}

Approved models create a gatekeeping layer in which the rule may bend, but only if justified to someone with recognized authority. 

\subsection{Escalated: High-Stakes Deviations}

Finally, some deviations trigger review beyond immediate guideline stewards. In these cases, breaking from established standards required broader institutional consultation, sometimes involving brand teams, leadership, or legal stakeholders.
Participants described situations in which changes required cross-team sign-off or alignment with higher-level systems.

\begin{formal}
``If a local reporter is writing something, \ldots their editor says, oh, this chart, like, you should check with the graphics team on this chart.''

\hfill
\palpha
\end{formal}

Escalation was most common when deviations affected brand identity, accessibility compliance, or public accountability. While not every exception required such review, high-visibility or high-risk cases did.

\subsection{Synthesis}
Importantly, flexibility was often intentional. Some teams explicitly resisted over-formalization to preserve creativity or responsiveness. The \acro{SITUATED FLEXIBILITY} dimension, therefore, reveals style guides as negotiated systems of constraint, where rigidity and flexibility are calibrated to organizational risk, audience, and capacity. Understanding these conditions clarifies what a rule says and how it lives in practice.\\
\par

\begin{figure*}
    \centering
    \includegraphics[width=\linewidth]{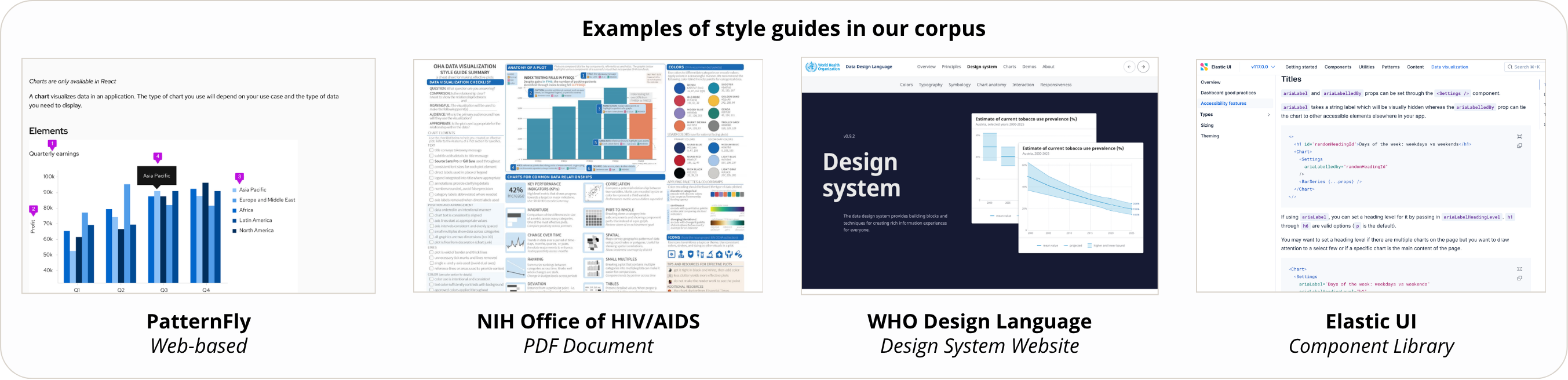}
    \caption{Representative visualization style guides analyzed in this study, illustrating the diversity of publication formats, including websites, PDF documents, design systems, and component libraries.}
    \label{fig:examples}
\end{figure*}

\section{The Partial Observability of Style Guides}

To illustrate the analytical value of \framework, we examined a corpus of publicly available visualization style guides. Our aim was not to fully reconstruct organizations' style-guide practices. Instead, we sought to identify which aspects of PRISM are directly observable from documentation and which are only partially observable.

The corpus consisted of 53 visualization style guides previously collected and analyzed by Ottley~\cite{ottley2026consensus}. The guides were originally identified through the Data Visualization Style Guide repository~\cite{dvsg2026}, a community-maintained collection spanning journalism, government, public health, nonprofit organizations, academic institutions, and technology companies. At the time of our analysis, 50 guides remained publicly accessible and published in English; three websites were no longer available and were therefore excluded. We provide representative examples in Figure~\ref{fig:examples}.

\subsection{Coding Procedure}

\new{Two researchers independently coded each style guide using the PRISM framework. During coding, we adopted an intentionally generous interpretation of public evidence in order to avoid underestimating the observable aspects of style-guide practice. For example, guides that embedded rules directly into software or reusable components were coded as exhibiting technological enforcement, even if enforcement was not explicitly discussed. Likewise, references directing readers to a community or support forum were treated as evidence of community-based governance. These inference rules were applied consistently throughout the analysis.}

\new{Following independent coding, the researchers jointly reviewed every disagreement alongside the recorded evidence and the original style-guide materials until consensus was reached. Most disagreements arose from one of two sources: (1) relevant information distributed across multiple webpages or linked documentation that one coder had initially overlooked, or (2) refinement of construct definitions during adjudication. The study corpus, organization-level metadata, and consensus coding annotations are available at \url{https://washuvis.github.io/prism}.}

\begin{figure}[b]
\centering
\begin{tikzpicture}
\footnotesize
\def\barheight{0.34}
\def\scale{0.055} 

\foreach \label/\value/\count/\y in {
    {\acro{rules}}/100/50/0,
    {\purpose}/52/26/-0.55,
    {\acro{inst. enforcers}}/38/19/-1.10,
    {\acro{sit. flexibility}}/26/13/-1.65
} {
    \node[anchor=east] at (0,\y) {\label};
    \draw[fill=gray!12, draw=none] (0.15,\y-\barheight/2) rectangle (5.65,\y+\barheight/2);
    \draw[fill=nytAccent!85, draw=none] (0.15,\y-\barheight/2) rectangle ({0.15+\value*\scale},\y+\barheight/2);
    \node[anchor=west] at (5.8,\y) {\count };
}


\end{tikzpicture}
\caption{\new{Observable \framework\ dimensions in 50 publicly available visualization style guides. Rules and mechanisms were visible in all guides, while the remaining dimensions were substantially less observable.}}
\label{fig:prism_observability}
\end{figure}
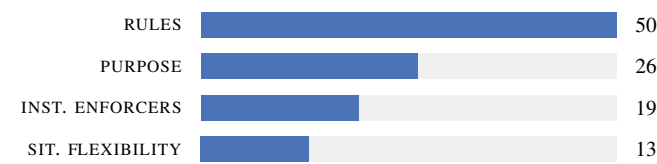

\subsection{What Is Visible: \rules}

Every analyzed style guide contained evidence of the \rules\ dimension. All 50 guides specified explicit standards (e.g., color palettes, typography rules, chart specifications, or accessibility requirements), while the majority ($n=47$) also included illustrative examples such as annotated charts, before--after comparisons, or ``do'' and ``don't'' examples. Eighteen provided reusable templates or components, including design tokens, starter files, or preconfigured chart styles.


Although the specific implementation varied across organizations, the emphasis on rules and mechanisms was remarkably consistent. These elements are readily documented, transferable, and amenable to formalization. Consequently, prior work has naturally focused on extracting and categorizing these visible components, treating style guides primarily as collections of prescriptive design rules~\cite{ottley2026consensus,elder2020should}.

\subsection{What Is Not Visible: Contextual Dimensions}

\new{In contrast, evidence for the remaining dimensions was substantially less common. Purpose was identifiable in only 26 organizations, institutional enforcers in 19, and situated flexibility in 13 (Fig.~\ref{fig:prism_observability})}.

\begin{itemize}

\item \textbf{\purpose.} Although guides frequently prescribe \emph{what} designers should do, they rarely explain \emph{why} those rules exist. Motivations such as accessibility, branding, editorial consistency, or production efficiency were often implied rather than explicitly articulated.

\item \textbf{\enforcers.} Fewer than half of the guides described how adherence is maintained. Some design systems implied technological enforcement through reusable components or automated tooling, but broader governance structures such as review processes, and organizational accountability were rarely documented.

\item \textbf{\flexibility.} Many guides employed authoritative language such as ``must'' or ``always,'' and only 13 explicitly stated that rules depend on context. None described formal escalation procedures, approval workflows, or circumstances under which guidance should be overridden. Our interview findings suggest that these decisions are instead negotiated through professional judgment and organizational norms.

\end{itemize}

\subsection{Implications}

The asymmetry observed across the corpus suggests that the partial observability of style guides is structural rather than incidental. Public-facing documents primarily communicate intervention mechanisms (\textit{what practitioners should do}) while providing substantially less insight into why those mechanisms exist, who maintains them, or when they should be adapted.

\framework\ makes this distinction explicit by separating the dimensions that are readily observable from documentation (\rules) from those that typically require organizational context (\purpose, \enforcers, and \flexibility). This distinction highlights the limitations of artifact-based analyses and motivates complementary methods, such as interviews, that can surface the organizational reasoning, governance structures, and contextual decision-making omitted from public documentation.

\section{Discussion: The Myths and Realities}

Our findings suggest that visualization style guides operate in ways that differ from how they are often discussed in research and practice. Instead of framing these differences as contradictions, we present them as five common assumptions that our interviews help clarify.

\medskip
\noindent
\textbf{Myth 1: Style Guides Are Primarily Best Practices} \hspace{1em}
A rule in a style guide rarely tells the whole story.
``Use direct labeling.''
``Limit pie slices.''
``Use a single bar color.''  
These statements appear universal. Yet our interviews reveal that such rules often respond to highly specific organizational constraints, such as multi-platform publishing, accessibility requirements for particular audiences, or the need to standardize production for non-experts. For instance, documented limitations on color and size might exist because the same data representation is used in both printed magazines and social media posts.

Still, style guides are often interpreted as compilations of empirically grounded design principles.
While perceptual research and professional norms certainly influence some rules, our interviews indicate that many guidelines are motivated by constraints and preferences. The justifications are, unfortunately, frequently absent. As a result, style guides appear transferable when they are not. Without understanding these contextual pressures, the underlying problems, rules can appear universal when they are in fact locally optimized.

\begin{quote}
\centering
\textit{Rules are best understood alongside the organizational conditions that produced them.}
\end{quote}

\noindent
\textbf{Myth 2: Rules in Style Guides Are Always Evidence-Driven} \hspace{1em}
Another striking pattern is that many rules are adopted because they appear in other style guides. Participants often consulted other existing style guides when authoring their own. In doing so, certain norms were inherited rather than independently justified. Debates central to visualization research, such as pie versus donut charts or axis truncation, sometimes entered guides through precedent, preference, or brand alignment. This does not imply that research is absent from practice. It reveals that style guides function as cultural artifacts as much as they do as research-informed artifacts, and may serve as stabilizing artifacts within communities. Some rules persist because they create recognizable visual identity or production efficiency, even when empirical evidence is ambiguous or evolving.

For researchers, this suggests an uncomfortable truth: evidence does not automatically travel into practice. If research is to influence style systems, it must integrate with brand logic, production workflows, and governance structures, not merely demonstrate perceptual superiority.

\begin{quote}
\centering
\textit{Understanding how evidence travels, or fails to travel, into institutional guidance is a critical area for research.}
\end{quote}

\medskip
\noindent
\textbf{Myth 3: A Style Guide Is a Document} \hspace{1em}
Many participants described \acro{PDF} guides as secondary or symbolic artifacts. Actionable guidance lived in templates, component systems, wrappers, tokens, review workflows, and training sessions. 
Thus, the traditional model of a downloadable ``style guide document'' is largely symbolic. Real enforcement and adoption occur through embedding defaults into tools and workflow integration. This suggests that style guides are better understood as infrastructural or design systems, not just standalone documents. 

\begin{quote}
\centering
\textit{Documentation captures intent; tools and workflow integration operationalize it.}
\end{quote}

\medskip
\noindent
\textbf{Myth 4: More Rules Reduce the Need for Judgment} \hspace{1em}
Across cases, participants emphasized that judgment never disappears. Even highly codified systems required contextual interpretation. Effective guides distinguished between non-negotiable requirements (e.g., accessibility, brand-critical elements) and areas where flexibility was appropriate.
Instead of striving for exhaustive rule coverage, some teams focused on clarifying when and how rules could bend.

\begin{quote}
\centering
\textit{Flexibility is not viewed as failure but as necessary and desired when applying guidance across diverse contexts.}
\end{quote}

\medskip
\noindent
\textbf{Myth 5: Style Guides Are About Aesthetics} \hspace{1em}
While visual consistency was a common motivator, style guides frequently emerged in response to expertise bottlenecks, uneven visualization literacy, and governance challenges. 
A recurring theme was that many visualization creators are not experts. Participants described low visualization literacy, repeated correction cycles, and the need for training and scaffolding. 
In this sense, style guides function as mechanisms for distributing expertise and stabilizing decision-making under time pressure.

This reframes style guides as pedagogical tools as much as governance artifacts. They compensate for uneven knowledge across organizations.
For the \acro{VIS} community, this suggests that research on literacy, scaffolding, and decision support is directly relevant to style systems. The question is not only what rules are correct, but how guidance supports non-experts in making defensible decisions.

\begin{quote}
\centering
\textit{Style guides are as much about training as they are about visual form.}
\end{quote}
\section{\new{Implications for Practitioners and Future Work}}

\new{Participants, especially those at smaller organizations who developed their first visualization style guide, recalled having little guidance on where to begin. In the absence of an organizing framework, style guides often emerge by borrowing rules from other organizations or accumulating recommendations over time. 
\framework\ can offer a different starting point. For example, instead of asking \textit{What rules should our style guide contain?}, organizations can begin by asking \textit{What organizational problem are we trying to solve?} Identifying the underlying purpose naturally informs the intervention mechanisms, governance structures, and flexibility required, allowing teams to focus on guidance that addresses their specific organizational needs and avoid adopting practices that may not apply to their context.}

\new{This perspective also opens several directions for future work. \framework\ could inform tools that support the authoring and evolution of visualization style guides, guide organizational audits, or scaffold workshops for developing new guidance. More broadly, future research can investigate whether style guides developed using a purpose-driven framework improve adoption, onboarding, governance, and long-term maintenance.}

\section{\new{Limitations}}

\new{This work has several limitations. First, our interview sample consisted of nine practitioners from four Western countries who collectively authored twenty-six visualization style guides. Although these participants represented diverse sectors, the framework may not capture practices used in other cultural contexts, academic settings, or organizations not represented in our sample.}

\new{Second, our artifact analysis was limited to publicly available English-language style guides. As our findings suggest, public documentation provides only a partial view of organizational practice and may omit internal governance processes, contextual guidance, and organizational knowledge.}

\new{Finally, \framework\ should be viewed as an interpretive analytic framework rather than a complete taxonomy of visualization style guides. Future work involving additional organizations and contexts may refine or extend the framework.}

\section{\new{Open Materials and Reproducibility}}

\new{To support transparency and reproducibility, we provide a companion website that documents the \framework\ framework, the study corpus, organization-level metadata, consensus coding annotations, and companion resources: \url{https://washuvis.github.io/prism}. We do not release interview transcripts because they contain sensitive organizational information that could reasonably enable participant re-identification.}

\section{Conclusion}

Through interviews with authors and maintainers across journalism, government, industry, and consulting, we show that visualization style guides are situated responses to organizational pressures. The resulting systems vary in structure and rigidity, reflecting tradeoffs among consistency, flexibility, authority, and accountability. We introduce \framework\ and offer a vocabulary for describing these systems beyond their surface prescriptions. This framing shifts attention from what rules say to why they exist, how they are enacted, who sustains them, and when they bend. In doing so, it reframes style guides as evolving sociotechnical infrastructures rather than static documents.

\acknowledgments{%
    We sincerely thank the practitioners who generously volunteered their time to participate in this study. Their openness, candor, and willingness to share their experiences made this work possible and provided valuable insights into visualization practice.
	This work is supported in part by the National Science Foundation under Grants Nos. 2142977 and 2330245, which support the Engineering Research Center for Carbon Utilization Redesign through Biomanufacturing-Empowered Decarbonization (CURB).%
}
\clearpage

\bibliographystyle{abbrv-doi-hyperref}

\bibliography{template}

@book{bertin1967semiologie,
  author = {Bertin, Jacques},
  title = {{{S{\'e}Miologie Graphique: Les Diagrammes, Les {R}{\'e}Seaux, Les Cartes}}},
  year = {1967},
  publisher = {Esri Press},
  doi = {10.3406/ahess.1967.421596}
}

@article{braun2019reflecting,
  author = {Braun, Virginia and Clarke, Victoria},
  title = {{{Reflecting on Reflexive Thematic Analysis}}},
  journal = {Qualitative research in sport, exercise and health},
  year = {2019},
  volume = {11},
  number = {4},
  pages = {589--597},
  publisher = {Taylor \& Francis},
  doi = {10.1080/2159676x.2019.1628806}
}

@article{braun2021one,
  author = {Braun, Virginia and Clarke, Victoria},
  title = {{{One Size Fits All? What Counts as Quality Practice in (reflexive) Thematic Analysis?}}},
  journal = {Qualitative research in psychology},
  year = {2021},
  volume = {18},
  number = {3},
  pages = {328--352},
  publisher = {Taylor \& Francis},
  doi = {10.4135/9781036232078.n9}
}

@incollection{braun2023doing,
  author = {Braun, Virginia and Clarke, Victoria and Hayfield, Nikki and Davey, Louise and Jenkinson, Elizabeth},
  title = {{{Doing Reflexive Thematic Analysis}}},
  booktitle = {{Supporting research in counselling and psychotherapy: Qualitative, quantitative, and mixed methods research}},
  year = {2023},
  pages = {19--38},
  publisher = {Springer},
  doi = {10.1007/978-3-031-17299-1_3470}
}

@book{brinton1919graphic,
  author = {Brinton, Willard Cope},
  title = {{{Graphic Methods for Presenting Facts}}},
  year = {1919},
  publisher = {Engineering magazine company}
}

@article{campbell2021reflexive,
  author = {Campbell, Karen A and Orr, Elizabeth and Durepos, Pamela and Nguyen, Linda and Li, Lin and Whitmore, Carly and Gehrke, Paige and Graham, Leslie and Jack, Susan M},
  title = {{{Reflexive Thematic Analysis for Applied Qualitative Health Research}}},
  journal = {The qualitative report},
  year = {2021},
  volume = {26},
  number = {6},
  pages = {2011--2028},
  doi = {10.46743/2160-3715/2021.5010}
}

@incollection{card2009information,
  author = {Card, Stuart},
  title = {{{Information Visualization}}},
  booktitle = {{Human-computer interaction}},
  year = {2009},
  pages = {199--234},
  publisher = {CRC press},
  doi = {10.1201/9781420088861.ch10}
}

@inproceedings{choi2021toward,
  author = {Choi, Jinhan and Oh, Changhoon and Suh, Bongwon and Kim, Nam Wook},
  title = {{{Toward a Unified Framework for Visualization Design Guidelines}}},
  booktitle = {{Extended Abstracts of the 2021 CHI Conference on Human Factors in Computing Systems}},
  year = {2021},
  pages = {1--7},
  doi = {https://doi.org/10.1145/3411763.3451702}
}

@article{cleveland1984graphical,
  author = {Cleveland, William S and McGill, Robert},
  title = {{{Graphical Perception: Theory, Experimentation, and Application to the Development of Graphical Methods}}},
  journal = {Journal of the American statistical association},
  year = {1984},
  volume = {79},
  number = {387},
  pages = {531--554},
  publisher = {Taylor \& Francis},
  doi = {10.1080/01621459.1984.10478080}
}

@inproceedings{correll2019ethical,
  author = {Correll, Michael},
  title = {{{Ethical Dimensions of Visualization Research}}},
  booktitle = {{Proceedings of the 2019 CHI conference on human factors in computing systems}},
  year = {2019},
  pages = {1--13},
  doi = {10.1145/3290605.3300418}
}

@incollection{diakopoulos2018ethics,
  author = {Diakopoulos, Nicholas},
  title = {{{Ethics in Data-Driven Visual Storytelling}}},
  booktitle = {{Data-Driven Storytelling}},
  year = {2018},
  pages = {233--248},
  publisher = {AK Peters/CRC Press},
  doi = {10.1201/9781315281575-10}
}

@article{diehl2020studying,
  author = {Diehl, Alexandra and Kraus, Matthias and Abdul-Rahman, Alfie and El-Assady, Mennatallah and Bach, Benjamin and Laramee, Robert Steven and Keim, Daniel and Chen, Min},
  title = {{{Studying Visualization Guidelines According to Grounded Theory}}},
  journal = {arXiv preprint arXiv:2010.09040},
  year = {2020},
  doi = {10.1111/cgf.13993}
}

@misc{dvsg2026,
  author = {Schwabish, Jonathan and Cesal, Amy and Wilson, Alan and Graze, Maxene},
  title = {{{Data Visualization Style Guidelines}}},
  journal = {Data Visualization Style Guidelines},
  year = {(accessed March 1, 2025)},
  howpublished = {https://www.datavizstyleguide.com/}
}

@article{elder2020should,
  author = {Elder, Kevin L and Cesal, Amy A},
  title = {{{Should We Teach Data Visualization Using Data Visualization Style Guides?}}},
  journal = {Issues in Information Systems},
  year = {2020},
  volume = {21},
  number = {4},
  pages = {264},
  doi = {10.1145/3377814.3381719}
}

@article{franconeri2021science,
  author = {Franconeri, Steven L and Padilla, Lace M and Shah, Priti and Zacks, Jeffrey M and Hullman, Jessica},
  title = {{{The Science of Visual Data Communication: What Works}}},
  journal = {Psychological Science in the public interest},
  year = {2021},
  volume = {22},
  number = {3},
  pages = {110--161},
  publisher = {Sage Publications Sage CA: Los Angeles, CA},
  doi = {10.1177/15291006211051956}
}

@article{friedman1996value,
  author = {Friedman, Batya},
  title = {{{Value-Sensitive Design}}},
  journal = {interactions},
  year = {1996},
  volume = {3},
  number = {6},
  pages = {16--23},
  publisher = {ACM New York, NY, USA},
  doi = {10.1145/242485.242493}
}

@article{friedman2017survey,
  author = {Friedman, Batya and Hendry, David G and Borning, Alan and others},
  title = {{{A Survey of Value Sensitive Design Methods}}},
  journal = {Foundations and Trends{\textregistered} in Human--Computer Interaction},
  year = {2017},
  volume = {11},
  number = {2},
  pages = {63--125},
  publisher = {Now Publishers, Inc.},
  doi = {10.1561/1100000015}
}

@article{hullman2011visualization,
  author = {Hullman, Jessica and Diakopoulos, Nick},
  title = {{{Visualization Rhetoric: Framing Effects in Narrative Visualization}}},
  journal = {IEEE transactions on visualization and computer graphics},
  year = {2011},
  volume = {17},
  number = {12},
  pages = {2231--2240},
  publisher = {IEEE},
  doi = {10.1109/tvcg.2011.255}
}

@article{kandogan2016grounded,
  author = {Kandogan, Eser and Lee, Hanseung},
  title = {{{A Grounded Theory Study on the Language of Data Visualization Principles and Guidelines}}},
  journal = {Electronic Imaging},
  year = {2016},
  volume = {28},
  pages = {1--9},
  publisher = {Society for Imaging Science and Technology},
  doi = {10.2352/issn.2470-1173.2016.16.hvei-132}
}

@inproceedings{lin2025makes,
  author = {Lin, Kylie and Ru, Sean Sheng-tse and Rapp, David N and Guan, Hui and Xiong Bearfield, Cindy},
  title = {{{What Makes a Visualization Visually Complex?}}},
  booktitle = {{Proceedings of the Extended Abstracts of the CHI Conference on Human Factors in Computing Systems}},
  year = {2025},
  pages = {1--7},
  doi = {10.1145/3706599.3719983}
}

@inproceedings{mckinley2025trustworthy,
  author = {McKinley, Oen G and Pandey, Saugat and Ottley, Alvitta},
  title = {{{Trustworthy by Design: the Viewer's Perspective on Trust in Data Visualization}}},
  booktitle = {{Proceedings of the 2025 CHI Conference on Human Factors in Computing Systems}},
  year = {2025},
  pages = {1--17},
  doi = {10.1145/3706598.3713824}
}

@article{moritz2018formalizing,
  author = {Moritz, Dominik and Wang, Chenglong and Nelson, Greg L and Lin, Halden and Smith, Adam M and Howe, Bill and Heer, Jeffrey},
  title = {{{Formalizing Visualization Design Knowledge as Constraints: Actionable and Extensible Models in Draco}}},
  journal = {IEEE transactions on visualization and computer graphics},
  year = {2018},
  volume = {25},
  number = {1},
  pages = {438--448},
  publisher = {IEEE},
  doi = {10.31219/osf.io/3eg9c}
}

@inproceedings{ottley2026consensus,
  author = {Ottley, Alvitta},
  title = {{{Consensus and Contradictions: a Cross-Organizational Analysis of Visualization Style Guides}}},
  booktitle = {{Proceedings of the 2026 CHI Conference on Human Factors in Computing Systems}},
  year = {2026},
  pages = {1--21},
  doi = {10.1145/3772318.3791628}
}

@inproceedings{pandey2023you,
  author = {Pandey, Saugat and McKinley, Oen G and Crouser, R Jordan and Ottley, Alvitta},
  title = {{{Do You Trust What You See? Toward a Multidimensional Measure of Trust in Visualization}}},
  booktitle = {{2023 IEEE Visualization and Visual Analytics (VIS)}},
  year = {2023},
  pages = {26--30},
  doi = {10.1109/vis54172.2023.00014},
  organization = {IEEE}
}

@article{satyanarayan2016vega,
  author = {Satyanarayan, Arvind and Moritz, Dominik and Wongsuphasawat, Kanit and Heer, Jeffrey},
  title = {{{Vega-Lite: a Grammar of Interactive Graphics}}},
  journal = {IEEE transactions on visualization and computer graphics},
  year = {2016},
  volume = {23},
  number = {1},
  pages = {341--350},
  publisher = {IEEE},
  doi = {10.32614/cran.package.vegalite}
}

@incollection{shneiderman2003eyes,
  author = {Shneiderman, Ben},
  title = {{{The Eyes Have It: a Task by Data Type Taxonomy for Information Visualizations}}},
  booktitle = {{The craft of information visualization}},
  year = {2003},
  pages = {364--371},
  publisher = {Elsevier},
  doi = {10.1016/b978-155860915-0/50046-9}
}

@inproceedings{wang2024card,
  author = {Wang, Zezhong and Hao, Shan and Carpendale, Sheelagh},
  title = {{{Card-Based Approach to Engage Exploring Ethics in {AI} For Data Visualization}}},
  booktitle = {{Extended Abstracts of the CHI Conference on Human Factors in Computing Systems}},
  year = {2024},
  pages = {1--7},
  doi = {10.1145/3613905.3650972}
}

@inproceedings{wongsuphasawat2016towards,
  author = {Wongsuphasawat, Kanit and Moritz, Dominik and Anand, Anushka and Mackinlay, Jock and Howe, Bill and Heer, Jeffrey},
  title = {{{Towards a General-Purpose Query Language for Visualization Recommendation}}},
  booktitle = {{Proceedings of the workshop on human-in-the-loop data analytics}},
  year = {2016},
  pages = {1--6},
  doi = {10.1145/2939502.2939506}
}

@inproceedings{zeng2023review,
  author = {Zeng, Zehua and Battle, Leilani},
  title = {{{A Review and Collation of Graphical Perception Knowledge for Visualization Recommendation}}},
  booktitle = {{Proceedings of the 2023 CHI conference on human factors in computing systems}},
  year = {2023},
  pages = {1--16},
  doi = {10.1145/3544548.3581349}
}

\appendix 

\clearpage
\section{Interview Protocol}
\label{app:interview-protocol}

We conducted semi-structured interviews using the following protocol. Questions were adapted as needed to reflect participants’ roles, organizational contexts, and specific style guides.

\begin{scriptsize}
\noindent
\textbf{Background and Role}\vspace{-.5em}
\begin{enumerate}[noitemsep]
    \item How many data visualization style guides have you authored? When was the first one created?
    \item Can you describe your role and how you became responsible for developing a data visualization style guide?
    \item What is your relationship to the style guide? (e.g., initiated it, contributed sections, maintain it, enforce it)
\end{enumerate}

\noindent
\textbf{Origins and Motivations}\vspace{-.5em}
\begin{enumerate}[noitemsep]
    \setcounter{enumi}{3}
    \item What prompted the need for a data visualization style guide in your organization or project? What problem were you trying to solve at the time?
    \item Who was the intended audience, and how did that shape your decisions?
    \item Who initiated the effort, and who else was involved in shaping the first version?
    \item Looking back, what would you say were the top two to three goals of the guide when it was first created?
\end{enumerate}

\noindent
\textbf{Content and Structure}\vspace{-.5em}
\begin{enumerate}[noitemsep]
    \setcounter{enumi}{7}
    \item How did you determine which sections to include (e.g., chart types, color, accessibility, interaction)?
    \item Were there any sections you discussed but ultimately did not include? 
    \item Can you walk me through how a specific guideline was written? How did that rule get proposed? Was there any debate or pushback? Did you consult research, examples, or external guides?
    
\end{enumerate}

\noindent
\textbf{Evidence, Values, and Trade-offs}\vspace{-.5em}
\begin{enumerate}[noitemsep]
    \setcounter{enumi}{10}
    \item When writing or revising guidelines, what kinds of evidence did you rely on? (e.g., research papers, industry norms, internal experiments, intuition)
    \item Did you explicitly discuss values such as clarity, accuracy, accessibility, brand consistency, or trust? How did these values shape particular rules?
    \item Can you recall a situation where two values conflicted (e.g., clarity vs.\ completeness, or brand vs.\ accessibility)? How was that conflict resolved in the guide?
\end{enumerate}

\textbf{Use, Adoption, and Maintenance}\vspace{-.5em}
\begin{enumerate}[noitemsep]
    \setcounter{enumi}{13}
    \item How do you expect people in your organization to use the style guide in their day-to-day work?
    \item How did you encourage organizational adoption of the guide?
    \item What tools, templates, or delivery formats did you choose, and why?
    \item How do you decide when to update or revise the style guide? What typically triggers revisions (e.g., incidents, new tools, regulations, rebranding)?
    \item Do you receive feedback or complaints about the guide? What kinds of issues tend to arise?
\end{enumerate}

\noindent
\textbf{Future Needs and Reflections}\vspace{-.5em}
\begin{enumerate}[noitemsep]
    \setcounter{enumi}{18}
    \item What were the biggest challenges you faced while creating the guide?
    \item Looking back, what worked well, and what would you change?
    \item Are there kinds of support that would make it easier to create or maintain a style guide? (e.g., templates, tooling, research syntheses, evaluation methods)
    \item Is there anything about the development of your style guide that we have not discussed but that you think is important for understanding how it works in your organization?
\end{enumerate}
\end{scriptsize}

\newpage
\section{Coverage Reports}
\label{app:participant-coverage}

Coverage of PRISM themes across interview participants.
Blue squares (\includegraphics[width=5.5pt]{imgs/yes.pdf}) indicate that a theme was discussed during an interview. Because interviews were conversational and participants occupied different organizational roles, the absence of a theme should not be interpreted as evidence that it was absent from an organization or unimportant in practice. We include this figure to illustrate the breadth of evidence supporting each dimension of the PRISM framework across participants.

\begin{figure}[h!]

\footnotesize

\resizebox{.4\textwidth}{!}{%
\renewcommand{\arraystretch}{0.9}
\setlength{\tabcolsep}{1pt}
\begin{tabular}{@{}>{\raggedright\arraybackslash}p{1.3cm}@{\hspace{6pt}}>{\raggedleft\arraybackslash}p{2.4cm}@{\hspace{6pt}}*{9}{>{\centering\arraybackslash}p{0.55cm}@{\hspace{2pt}}}@{}}

& & \rotatebox{45}{ALPHA} & \rotatebox{45}{BETA} & \rotatebox{45}{DELTA} & \rotatebox{45}{EPSILON} & \rotatebox{45}{ETA} & \rotatebox{45}{GAMMA} & \rotatebox{45}{IOTA} & \rotatebox{45}{THETA} & \rotatebox{45}{ZETA} \\ 
\multirow{6}{*}{\parbox{1cm}{\centering\textsc{Purpose}}} & \textbf{Consistency} & \yes & \yes & \yes & \yes & \yes & \yes & \yes & \yes & \yes \\
 & \textbf{Alignment} & \yes & \yes & \yes & \yes & \yes & \yes & \yes & \yes & \yes \\
 & \textbf{Capacity} & \yes & \yes & \yes & \yes & \yes & \yes & \yes & \yes & \yes \\
 & \textbf{Update} & \yes & \no & \yes & \yes & \yes & \yes & \yes & \yes & \yes \\
 & \textbf{Constraints} & \yes & \yes & \yes & \yes & \yes & \yes & \yes & \yes & \yes \\
 & \textbf{Accountability} & \no & \no & \yes & \yes & \no & \yes & \yes & \yes & \no \\ \\ \\

\multirow{6}{*}{\parbox{1.3cm}{\centering\textsc{Rules \&\\Mechanisms}}} & \textbf{Audit} & \yes & \yes & \yes & \yes & \yes & \yes & \yes & \yes & \yes \\
 & \textbf{Standards} & \yes & \yes & \yes & \yes & \yes & \yes & \yes & \yes & \yes \\
 & \textbf{Examples} & \yes & \yes & \yes & \yes & \yes & \yes & \yes & \yes & \yes \\
 & \textbf{Templates} & \yes & \yes & \yes & \yes & \no & \yes & \no & \yes & \yes \\
 & \textbf{Automation} & \yes & \yes & \yes & \yes & \no & \yes & \yes & \yes & \yes \\
 & \textbf{Training} & \yes & \yes & \yes & \yes & \yes & \yes & \yes & \yes & \yes \\ \\ \\

\multirow{4}{*}{\parbox{1.3cm}{\centering\textsc{Institutional\\Enforcers}}} & \textbf{Individual} & \yes & \yes & \yes & \yes & \yes & \yes & \yes & \yes & \yes \\
 & \textbf{Community} & \no & \yes & \yes & \yes & \yes & \yes & \no & \yes & \yes \\
 & \textbf{Workflow} & \yes & \yes & \yes & \yes & \yes & \yes & \yes & \yes & \yes \\
 & \textbf{Technology} & \yes & \no & \no & \yes & \no & \yes & \no & \yes & \yes \\ \\ \\

\multirow{4}{*}{\parbox{1.3cm}{\centering\textsc{Situated\\Flexibility}}} & \textbf{Strict} & \yes & \yes & \yes & \yes & \no & \yes & \yes & \yes & \no \\
 & \textbf{Contextual} & \yes & \yes & \no & \yes & \yes & \yes & \yes & \yes & \yes \\
 & \textbf{Approved} & \yes & \yes & \no & \no & \no & \no & \yes & \no & \no \\
 & \textbf{Escalated} & \yes & \no & \no & \no & \no & \no & \no & \no & \no \\
\end{tabular}%
}

\centering
\label{fig:participant_coverage_matrix}
\end{figure}

\end{document}